\documentclass[letter,twocolumn]{jpsj3}

\title{
  Variational Wavefunction for the Periodic Anderson Model\\
  with Onsite Correlation Factors
}

\author{Katsunori Kubo and Hiroaki Onishi}
\inst{
Advanced Science Research Center, Japan Atomic Energy Agency,
Tokai, Ibaraki 319-1195, Japan}

\abst{
We propose a variational wavefunction
containing parameters to tune the probabilities of
all the possible onsite configurations for the periodic Anderson model.
We call it the full onsite-correlation wavefunction (FOWF).
This is a simple extension of the Gutzwiller wavefunction (GWF),
in which one parameter is included to tune the double occupancy
of the $f$ electrons at the same site.
We compare the energy of the GWF and the FOWF
evaluated by the variational Monte Carlo method
and that obtained with the density-matrix renormalization group method.
We find that the energy is considerably improved in the FOWF.
On the other hand,
the physical quantities do not change significantly
between these two wavefunctions
as long as they describe the same phase, such as the paramagnetic phase.
From these results,
we not only demonstrate the improvement by the FOWF,
but we also gain insights on the applicability and limitation
of the GWF to the periodic Anderson model.
}

\begin{document}
\maketitle

In solid states,
the Coulomb interaction between electrons
reduces or renormalizes the energy scale for the electrons.
An extreme example of such an electron correlation effect
is the heavy-fermion phenomenon.
In heavy-fermion systems,
the effective mass of the electrons,
which is proportional to the inverse of the energy scale,
can become a thousand times larger than the bare electron mass.
Then, exotic order such as unconventional superconductivity has been found,
in particular, in the vicinity of the quantum critical points
of heavy-fermion materials.

As a canonical model for the heavy-fermion phenomenon,
the periodic Anderson model has been employed,
which is composed of conduction and $f$ electrons,
and the effects of the Coulomb interaction between $f$ electrons
have been studied by various methods.
In particular,
theories based on a variational wavefunction
called the Gutzwiller wavefunction (GWF)
have succeeded in describing
the heavy-fermion state.~\cite{Rice1985,Shiba1986,Rice1986,Fazekas1987}
In the GWF,
one parameter, called the Gutzwiller parameter,
is introduced to tune the probability of the double occupancy
of $f$ electrons on the same site.
The above theories assumed the paramagnetic (PM) state;
however, within the GWF,
it has also been revealed that
the PM state is unstable against magnetic order
in the parameter region
where the heavy-fermion state is realized.~\cite{Rice1985,Rice1986,
  Reynolds1992,Dorin1993JAP,Dorin1993PRB,
  Yokoyama1997,Watanabe2009,Kubo2013PRB,Kubo2015}

To overcome this difficulty,
we should improve the variational wavefunction.
For this purpose,
it is useful to consult the literature on other multiorbital systems
since the periodic Anderson model
is also a multiorbital system
with the conduction band and $f$ orbital.
For multiorbital Hubbard models,
parameters that tune the probabilities
of all the possible onsite configurations
have been introduced.~\cite{Okabe1997, Bunemann1998}
Here, we call this type of wavefunction
the full onsite-correlation wavefunction (FOWF).
The variational Monte Carlo method has been applied to the FOWF
to evaluate the energy and physical quantities
in the multiorbital Hubbard models.~\cite{Kobayashi2006,Kubo2009,Kubo2011}

It is natural to use the FOWF for the multiorbital Hubbard models
since all the orbitals are subjected to the Coulomb interactions.
On the other hand, in the periodic Anderson model,
it seems to be sufficient to consider the GWF at first glance
since the Coulomb interaction acts only on the electrons in the $f$ orbital.
However, the antiferromagnetic correlation
between the conduction and $f$ electrons is also important
in the periodic Anderson model.
Actually, it results in the Kondo phenomena
and the Ruderman--Kittel--Kasuya--Yosida interaction,
which have been central issues in condensed matter physics.
Thus, it would be a natural improvement
to consider all the possible onsite correlations,
including the spin-dependent correlations
between the conduction and $f$ electrons,
in addition to the $f$-$f$ correlation.

In this paper,
we consider the FOWF for the periodic Anderson model
and apply the variational Monte Carlo method.~\cite{Shiba1986}
As a benchmark of the wavefunction,
we study the model on a one-dimensional chain
and compare the energy with that obtained
with the density-matrix renormalization group (DMRG) method.~\cite{White1992}
Then, we discuss how much the energy is improved from the GWF.
We also compare physical quantities
evaluated by the GWF and by the FOWF.
In this paper,
we not only emphasize the improvement by the FOWF,
but also note the situations where the GWF is still useful
by comparing the results between these two wavefunctions.

The periodic Anderson model is given by
\begin{equation}
  \begin{split}
    H=&\sum_{k \sigma}\epsilon_{k}
    c^{\dagger}_{k \sigma}c_{k \sigma}
    +\sum_{i \sigma}\epsilon_f n_{f i \sigma}\\
    -&V\sum_{k \sigma}(f^{\dagger}_{k \sigma}c_{k \sigma}
                            +c^{\dagger}_{k \sigma}f_{k \sigma})
    +U\sum_{i}n_{f i \uparrow}n_{f i \downarrow},
  \end{split}
\end{equation}
where
$c^{\dagger}_{k \sigma}$
and
$f^{\dagger}_{k \sigma}$
are the creation operators of the conduction and $f$ electrons,
respectively, with momentum $k$ and spin $\sigma$,
and $n_{f i \sigma}$ is the number operator
of the $f$ electron with spin $\sigma$ at site $i$.
$\epsilon_{k}$ is the kinetic energy of the conduction electron,
$\epsilon_f$ is the energy level of the $f$ orbital,
$V$ is the hybridization matrix element
between the conduction and $f$ electrons,
and $U$ is the onsite Coulomb interaction for the $f$ orbital.
The Coulomb interaction $U$ is large
since the spatial extent of the $f$-orbital wavefunction is narrow.
In this study, we simply take $U \rightarrow \infty$
since we will obtain qualitatively the same results
even for a finite $U$ as long as it is sufficiently large,
for example, comparable to the band width of the conduction electrons,
to describe $f$-electron compounds.
Actually, within the GWF, we have obtained qualitatively similar results
for a finite $U$.~\cite{Kubo2015}
For the kinetic energy of the conduction electrons,
we consider only the nearest-neighbor hopping
on a one-dimensional chain,
and it is given by $\epsilon_{k}=-2t \cos k$,
where $t$ is the hopping integral and we set the lattice constant as unity.

We apply the variational Monte Carlo method to the model.~\cite{Shiba1986}
We consider two types of wavefunction
as mentioned in the introductory part.

One is the GWF,
which has been used to study this model,
for example, by the variational Monte Carlo method
as in this study.~\cite{Shiba1986,Yokoyama1997,Watanabe2009,Kubo2015}
For $U \rightarrow \infty$, the Gutzwiller parameter is zero,
and the GWF is given by
\begin{equation}
  | \psi_{\text{G}} \rangle
  =P_{\text{G}} | \phi \rangle,
\end{equation}
where
\begin{equation}
  P_{\text{G}}=\prod_{i}[1-n_{f i \uparrow}n_{f i \downarrow}].
\end{equation}
The projection operator $P_{\text{G}}$ excludes
the double occupancy of the $f$ electrons at the same site.
$| \phi \rangle$ is the one-electron part of the wavefunction.
We define it as the ground state
of a mean-field-type effective Hamiltonian,
which we will give later.

The other variational wavefunction is the FOWF.
It is defined as
\begin{equation}
  | \psi_{\text{FO}} \rangle
  =P_{\text{FO}} | \phi \rangle,
\end{equation}
where
\begin{equation}
  P_{\text{FO}}=\prod_{\gamma i}[1-(1-g_{\gamma})P_{\gamma i}],
\end{equation}
with $P_{\gamma i}=|\gamma i \rangle \langle \gamma i|$,
which is the projection operator to state $\gamma$ at site $i$.
$|\phi\rangle$ is the one-electron part as in the GWF.
While this type of wavefunction has been used
to study multiorbital Hubbard models,~\cite{Okabe1997,Bunemann1998,
  Kobayashi2006,Kubo2009,Kubo2011}
it has not been applied to the periodic Anderson model.
Since we consider one conduction band and one $f$ orbital,
there are 16 onsite states.
By considering symmetry and conservation of the number of electrons
for each spin,
we can reduce the number of independent variational parameters
that we have to optimize.
In addition,
for $\gamma$ denoting a configuration with doubly occupied $f$ electrons,
$g_{\gamma}=0$ since $U \rightarrow \infty$.
The FOWF reduces to the GWF
if we set $g_{\gamma}=1$ for configurations $\gamma$
without doubly occupied $f$ electrons.

In this study, we consider
the PM and ferromagnetic (FM) states, i.e., spatially uniform states.
Then, the effective Hamiltonian is given by~\cite{Kubo2015}
\begin{equation}
  H_{\text{eff}}=\sum_{k \sigma}
  (c^{\dagger}_{k \sigma} \ f^{\dagger}_{k \sigma})
  \begin{pmatrix}
    \epsilon_{k} & -\tilde{V}_{\sigma} \\[1.5ex]
    -\tilde{V}_{\sigma} & \tilde{\epsilon}_{f \sigma}
  \end{pmatrix}
  \begin{pmatrix}
    c_{k \sigma} \\[1.5ex]
    f_{k \sigma}
  \end{pmatrix},
\end{equation}
where $\tilde{V}_{\sigma}$ is the effective hybridization matrix element
and $\tilde{\epsilon}_{f \sigma}$ is the effective $f$-level,
which are also variational parameters.
For the PM state, they do not depend on the spin $\sigma$.
We diagonalize $H_{\text{eff}}$ and construct
its ground state $|\phi\rangle$
while fixing the electron number $n_{\sigma}$ per site of each spin $\sigma$.
In the PM state, $n_{\uparrow}=n_{\downarrow}$.
For the FM state,
the magnetization $M=n_{\uparrow}-n_{\downarrow}$ is a parameter
characterizing the state.

For each state, we evaluate the energy by the Monte Carlo method
and optimize the variational parameters that minimize the energy.
For the FM state, we also have to optimize the magnetization.
Then, we compare the energies of these states
with the same electron density $n=n_{\uparrow}+n_{\downarrow}$
and determine the ground state.
Other physical quantities can also be calculated by the Monte Carlo method
with the optimized variational parameters.

In this study, we set $U \rightarrow \infty$, $V=t$, and $n=1.25$.
The calculations are carried out for a 40-site chain
with a periodic boundary condition.

\begin{figure}
  \includegraphics[width=0.99\linewidth]
  {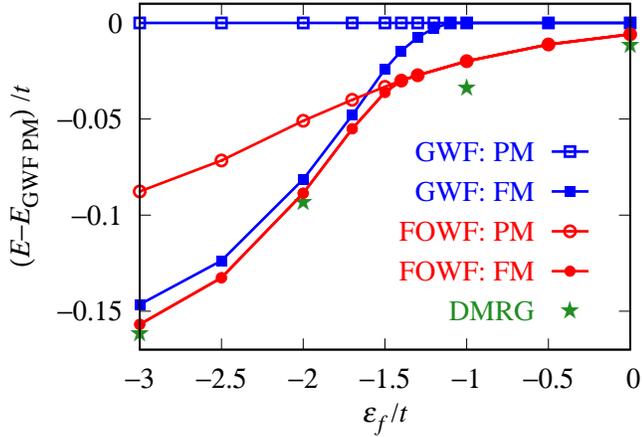}
  \caption{\label{E_n1.25}
    (Color online)
    Energy as functions of $\epsilon_f$
    measured from that of the Gutzwiller wavefunction (GWF)
    for the paramagnetic (PM) state $E_{\text{GWF PM}}$:
    the GWF (squares)
    and the full onsite-correlation wavefunction (FOWF, circles)
    for the PM state (open symbols)
    and for the ferromagnetic (FM) state (solid symbols).
    $U/t \rightarrow \infty$, $V/t=1$, and $n=1.25$.
    The energy obtained by the DMRG method
    for a sufficiently large value of the Coulomb interaction, $U/t=1000$,
    is also shown (stars).
  }
\end{figure}

Figure~\ref{E_n1.25} shows the energy $E$ per site
of the GWF and FOWF for the PM and FM states
as functions of $\epsilon_f$.
The energy is measured from that of the GWF for the PM state $E_{\text{GWF PM}}$.
The statistical errors are much smaller than the symbol sizes,
and we do not draw them.
In addition, we have also evaluated the energy
by the DMRG method~\cite{White1992} (stars in Fig.~\ref{E_n1.25}),
by which we can accurately determine the ground state
of a one-dimensional system as the present model
in an unbiased manner.
In the DMRG calculations,
we have performed extrapolation to zero truncation error
by using data for up to 800 DMRG states.

For both types of variational wavefunction,
the energy for the PM and FM states coincides
at higher values of $\epsilon_f$,
that is, the energy takes a minimum at $M=0$
and the system is in the PM phase.
At lower values of $\epsilon_f$,
the energy of the FM state becomes lower than that in the PM state,
and the ground state is FM there.
By the DMRG method,
we obtained the PM state for $\epsilon_f/t=0$ and $-1$
and the FM state for $\epsilon_f/t=-2$ and $-3$.

In both the PM and FM phases,
by improving the wavefunction from the GWF to the FOWF,
the energy becomes very close to that of the DMRG method.
In other words,
by only including the onsite projection factors,
we can greatly improve the energy.
The FM transition point,
where the FM energy departs from the PM energy,
shifts to a lower value of $\epsilon_f$,
since 
more fluctuations are included in the FOWF in comparison with in the GWF,
and the FM ordered state becomes unstable to a certain extent.

In principle,
the order of the transition can be determined by the slope of $E$:
it is second-order when the slope is continuous at the transition point
and first-order when the slope is discontinuous.
The order of the transition seems to be second-order
in both the wavefunctions,
while it is difficult to exclude
the possibility of a weak first-order transition
from the present numerical calculations on a finite-size lattice.

\begin{figure}
  \includegraphics[width=0.99\linewidth]
  {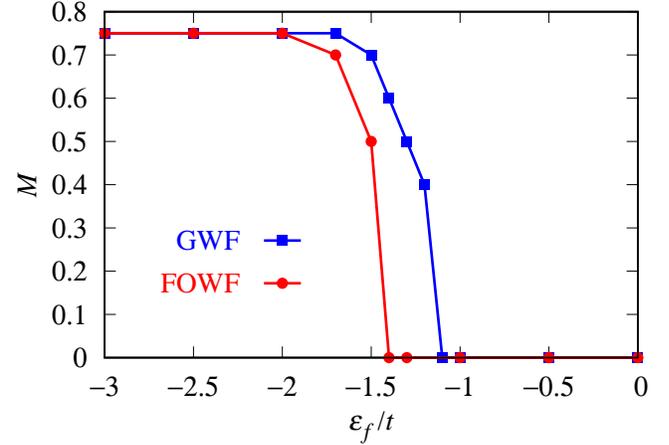}
  \caption{\label{magnetization_n1.25}
    (Color online)
    Magnetization for the GWF (squares) and FOWF (circles)
    as functions of $\epsilon_f$.
    $U/t \rightarrow \infty$, $V/t=1$, and $n=1.25$.
  }
\end{figure}

In Fig.~\ref{magnetization_n1.25}, we show the FM moment
as a function of $\epsilon_f$.
Note that the state with $M=2-n=0.75$ is a half-metallic state,
in which the Fermi surface for the up-spin electrons disappears.
The magnetization seems to develop from zero continuously,
which is consistent with the second-order transition
suggested from the behavior of the energy.
However, we mention again that
we should investigate larger lattices carefully
to determine the order of the transition
since we have to deal with states with smaller values of magnetization,
which require higher resolution.
This is beyond the scope of the present study.
The overall behavior of the magnetization is similar between the GWF and FOWF,
but the FM transition point shifts to a lower value of $\epsilon_f$
in the FOWF as discussed above.

\begin{figure}
  \includegraphics[width=0.99\linewidth]
  {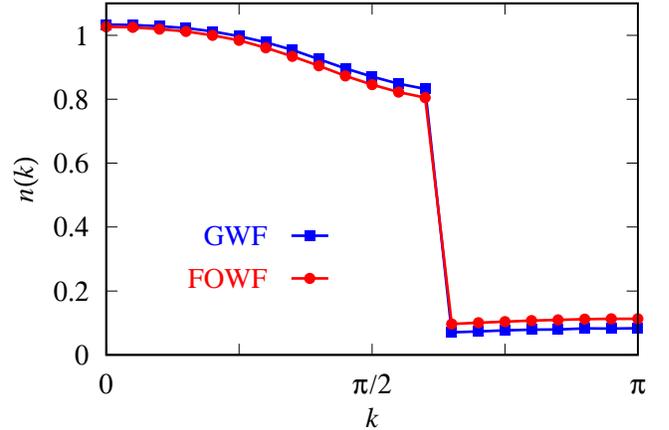}
  \caption{\label{nk_n1.25_efm1}
    (Color online)
    Momentum distribution functions
    in the PM phase
    of the GWF (squares)
    and
    FOWF (circles).
    $U/t \rightarrow \infty$, $V/t=1$, $n=1.25$, and $\epsilon_f/t=-1$.
  }
\end{figure}

\begin{figure}
  \includegraphics[width=0.99\linewidth]
  {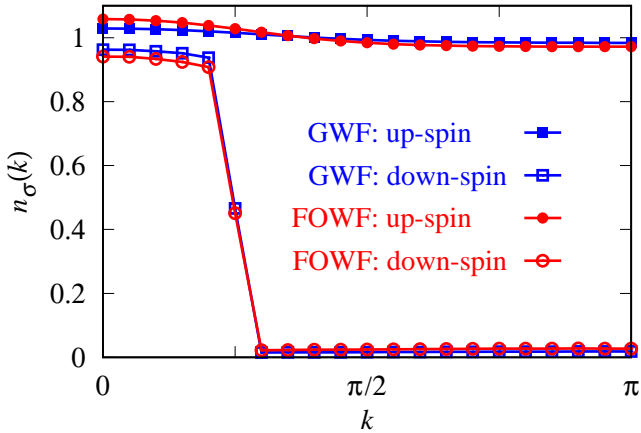}
  \caption{\label{nk_n1.25_efm2}
    (Color online)
    Momentum distribution functions
    in the FM phase
    of the GWF (squares)
    and
    FOWF (circles).
    Solid (open) symbols are for up-spin (down-spin) states.
    $U/t \rightarrow \infty$, $V/t=1$, $n=1.25$, and $\epsilon_f/t=-2$.
  }
\end{figure}

To investigate the electronic state further,
we also calculate the momentum distribution function.
The momentum distribution function is defined as
\begin{equation}
  n_{\sigma}(k)=
  \langle c^{\dagger}_{k \sigma}c_{k \sigma} \rangle
  + \langle f^{\dagger}_{k \sigma}f_{k \sigma} \rangle,
\end{equation}
where $\langle \cdots \rangle$ denotes the expectation value.
It does not depend on the spin $\sigma$ in the PM state:
$n_{\uparrow}(k)=n_{\downarrow}(k)=n(k)$.
The momentum distribution functions
of the GWF and FOWF are shown in Fig.~\ref{nk_n1.25_efm1} for $\epsilon_f/t=-1$
and in Fig.~\ref{nk_n1.25_efm2} for $\epsilon_f/t=-2$.
Both wavefunctions result
in the PM phase for $\epsilon_f/t=-1$
and
in the FM phase for $\epsilon_f/t=-2$
(see Figs.~\ref{E_n1.25} and \ref{magnetization_n1.25}).
The slight but finite increase in $n(k)$ in Fig.~\ref{nk_n1.25_efm1}
with $k$ above the Fermi momentum
is an artifact of the present wavefunctions
without intersite correlations.~\cite{Shiba1986,Yokoyama1987,Yokoyama1990}
$n_{\downarrow}(k)$ in Fig.~\ref{nk_n1.25_efm2}
is also an increasing function above the Fermi momentum for the same reason,
although it is invisible on this scale.

While the energy is significantly improved from the GWF to the FOWF,
the momentum distribution functions are very similar
between these wavefunctions.
Thus, we expect that
physical quantities can be evaluated accurately even by the GWF
as long as both wavefunctions describe the same phase.

In one-dimensional correlated systems,
we expect a Tomonaga--Luttinger liquid state in general.
However, we see a clear jump in the momentum distribution function
at the Fermi momentum in Figs.~\ref{nk_n1.25_efm1} and \ref{nk_n1.25_efm2},
indicating a Fermi liquid.
This is due to the fact that
we have used variational wavefunctions for Fermi liquid states.
To describe a Tomonaga--Luttinger liquid state,
we should carefully consider intersite correlations,
which are not included here.~\cite{Hellberg1991,Kawakami1992,Hellberg1992}
In addition,
the disappearance of the jump may be difficult
to observe for a finite-size lattice as in this study,
even if we improve the wavefunction further.

However, the size of the apparent jump may be used
to measure the correlation effect,
and conclusions extracted from this quantity
should be applicable to higher-dimensional systems,
where Fermi liquid states are expected.
In other words,
we expect that the characteristics of the variational wavefunctions
considered here will not depend strongly on the dimensionality,
although the applicability will depend strongly.
If the correlation effect becomes strong,
the size of the jump should be reduced.
Actually, the jump is reduced slightly in the FOWF
in comparison with in the GWF.

\begin{figure}
  \includegraphics[width=0.99\linewidth]
  {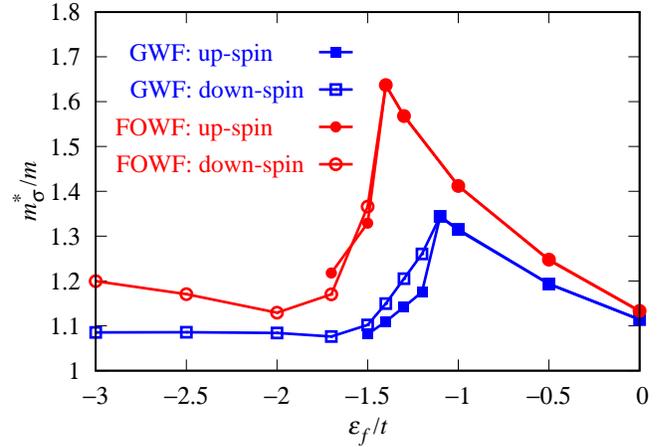}
  \caption{\label{effective_mass_n1.25}
    (Color online)
    Effective mass as functions of $\epsilon_f$
    of the GWF
    for up-spin (solid squares) and down-spin (open squares) electrons
    and
    of the FOWF 
    for up-spin (solid circles) and down-spin (open circles) electrons.
    $U/t \rightarrow \infty$, $V/t=1$, and $n=1.25$.
  }
\end{figure}

By using the jump $\Delta n_{\sigma}(k_{\text{F}})$ at the Fermi momentum,
we can define the effective mass $m^*_{\sigma}$ as
\begin{equation}
  \frac{m^*_{\sigma}}{m}=\frac{1}{\Delta n_{\sigma}(k_{\text{F}})},
\end{equation}
where $m$ is the bare electron mass.
Figure~\ref{effective_mass_n1.25} shows
the $\epsilon_f$ dependence of the effective mass.
Note that in the half-metallic state,
we cannot define the effective mass for up-spin electrons
since the Fermi surface disappears for them.
While the effective mass is enhanced in the FOWF as expected,
the overall behavior is similar between these wavefunctions.
The main difference is the position of the FM transition.
The PM phase extends to lower values of $\epsilon_f$ in the FOWF
and, as a result, the FOWF can attain a larger effective mass.
If we assume a PM state,
we can obtain a large effective mass even in the GWF,
and actually the heavy-fermion state has been discussed by using the GWF
with this assumption.~\cite{Rice1985,Shiba1986,Rice1986,Fazekas1987}
The present study provides a justification of this assumption
of the PM state to some extent.

However, the obtained effective mass in the FOWF is still not very large.
We have also evaluated the effective mass for $n=1.75$
since the correlation effects become stronger near half-filling, $n=2$.
Actually, the effective mass at the same $\epsilon_f$
in the PM phase becomes larger,
but the FM transition point shifts to a higher value of $\epsilon_f \simeq 1$
and we also do not obtain a very large effective mass in the PM phase
for $n=1.75$.
Thus, to obtain a larger effective mass,
we should further improve the wavefunction and/or revise the model
if we take the possibility of magnetic order into consideration.

Concerning other quantities,
such as the effective $f$-level $\tilde{\epsilon}_{f\sigma}$,
the energy gain at the transition,
and the contribution of the $f$ electrons to $\Delta n_{\sigma}(k_{\text{F}})$,
the behaviors do not change significantly from those in the GWF,
as the physical quantities explicitly presented in this paper.
See Refs.~\citen{Kubo2013PRB} and \citen{Kubo2015}
for details of these quantities.

To summarize,
we have proposed a wavefunction for the periodic Anderson model
containing parameters
to tune all the possible onsite configurations (we named it the FOWF)
to improve the Gutzwiller wavefunction (GWF).
Although the FOWF does not require
a large computational effort
since only the onsite variational parameters are included,
the energy is considerably improved
and becomes close to the value
obtained with the density-matrix renormalization group method.
However,
physical quantities, such as the magnetization and the effective mass,
do not change significantly between the GWF and FOWF
as long as they are in the same phase.
The main differences between these two wavefunctions
appear in the energy and in the position of the FM transition point.

From these observations,
we conclude that
we can use the GWF to discuss the heavy-fermion state
by assuming the PM state,
since we can expect that the PM region will become wider
when we improve the wavefunction,
but the physical quantities can be evaluated accurately even by the GWF.
However,
to discuss the magnetic transition point itself
and the behavior around it, such as quantum critical phenomena,
it is better to use a wavefunction beyond the GWF.

\begin{acknowledgments}
This work was supported by JSPS KAKENHI Grant Numbers
JP15K05191 
and
JP16K05494. 
Part of the computations were carried out
on the supercomputers at the Japan Atomic Energy Agency
and the Institute for Solid State Physics, the University of Tokyo.
\end{acknowledgments}



\end{document}